\documentclass[journal,transmag]{IEEEtran}
\usepackage{tikz}
\usetikzlibrary{decorations.pathmorphing} 
\usetikzlibrary{shapes.geometric}         
\usetikzlibrary{positioning}              
\usetikzlibrary{arrows.meta}              

\usepackage{enumitem}
\usepackage{hyperref}

\ifCLASSINFOpdf
\else
\fi
\begin{document}
%
\title{Towards Log Analysis with AI Agents: Cowrie Case Study}


\author{\IEEEauthorblockN{Enis Karaarslan\IEEEauthorrefmark{1},
Esin Güler\IEEEauthorrefmark{1},
Efe Emir Yüce\IEEEauthorrefmark{1} and
Çağatay Çoban\IEEEauthorrefmark{2}}
\IEEEauthorblockA{\IEEEauthorrefmark{1}MSKÜ Engineering Faculty, Department of Computer Engineering, Türkiye}
\IEEEauthorblockA{\IEEEauthorrefmark{2} MSKÜ Institute of Science, Deparment of AI}
\thanks{Corresponding author: E. Karaarslan (email: enis.karaarslan@mu.edu.tr}}



\IEEEtitleabstractindextext{%
\begin{abstract}
The scarcity of real-world attack data significantly hinders progress in cybersecurity research and education. Although honeypots like Cowrie effectively collect live threat intelligence, they generate overwhelming volumes of unstructured and heterogeneous logs, rendering manual analysis impractical. As a first step in our project on secure and efficient AI automation, this study explores the use of AI agents for automated log analysis. We present a lightweight and automated approach to process Cowrie honeypot logs. Our approach leverages AI agents to intelligently parse, summarize, and extract insights from raw data, while also considering the security implications of deploying such an autonomous system. Preliminary results demonstrate the pipeline's effectiveness in reducing manual effort and identifying attack patterns, paving the way for more advanced autonomous cybersecurity analysis in future work.
\end{abstract}

}

\maketitle

\IEEEdisplaynontitleabstractindextext

%
\IEEEpeerreviewmaketitle

\section{Introduction}
The ever-evolving landscape of cyber threats necessitates continuous study and analysis. However, for independent researchers, students, and small organizations, accessing real-world attack datasets is a significant hurdle, as this data is often proprietary. Honeypots provide an effective method for collecting live data by acting as decoy systems designed to be probed and attacked. While successful in data collection, a honeypot can generate hundreds of thousands of log entries daily, creating a secondary problem of data overload. Manually analyzing this data to identify patterns and attacker TTPs (Tactics, Techniques, and Procedures) is a time-consuming and inefficient process.

This study addresses the need for an accessible and automated method to analyze honeypot data. The primary contribution of this work is the design and implementation of a complete approach which uses AI agents. This system automates the entire workflow, from reading raw Cowrie logs to generating high-level summaries and visualizations. The goal is to provide a clear and extensible framework that can empower researchers to quickly gain insights from their honeypot data without requiring a heavy and expensive SIEM (Security Information and Event Management) infrastructure. AI agents are used as a solution.

In the next section, the fundamental concepts of honeypots and AI agents will be given. Then the related works will be given in Section 3. In Section 4, details of the system proposal are given. The implementation of the system is given in Section 5. Results are given and analyzed in Section 6. The conclusion and future works will be given in the last section.

\section{Fundamentals}
\subsection {Honeypots}

A honeypot is a security mechanism intended to lure and trap cyber attackers by mimicking a legitimate target system. Honeypots are artificial decoy systems that offer significant advantages in cybersecurity. Key advantages include providing real-time threat data, capturing only malicious traffic with zero false positives, revealing attackers' tactics, techniques, and procedures (TTP), and being used to test IDS/IPS systems. However, if configured incorrectly, they can pose a threat to the actual system and also pose legal risks, such as the attacker claiming to be a victim themselves. Importantly, honeypots are never a sufficient solution on their own and must be used in integration with other security systems.

Cowrie is a popular medium-interaction SSH and Telnet honeypot designed to log brute-force attacks and, more importantly, the shell interactions an attacker performs after gaining access. It outputs these interactions into structured JSON files, which serve as the raw data source for our project.

\subsection {AI agents}

An AI agent is an autonomous system designed to perform tasks by generating and executing its own action plans, leveraging a set of tools to achieve its objectives \cite{c4}. It leverages advanced natural language processing capabilities to comprehend complex goals, break them down into logical, sequential steps, and determine the necessary external tools and when to call them \cite{c5}. While conventional AI systems typically operate within predefined parameters and require explicit instructions for each task, AI agents demonstrate greater autonomy in goal-directed behavior \cite{c6}.

Recent progress in large language models (LLMs), with their enhanced reasoning capabilities, has significantly advanced the development of AI agents. Today's agents use these powerful LLMs as their central brain, enhancing them with additional components for memory, planning, using tools, and interacting with their environment \cite{c4}. This difference enables us to accomplish more complex tasks, such as implementing highly adaptive and intelligent honeypot systems specifically designed for strategic cyber deception and advanced threat intelligence gathering.

\subsection{Rule-Based Analysis}

A rule-based system is a simple yet effective form of an expert system that uses a collection of predefined rules to make decisions. In our context, the "expert" is a security analyst, and the rules are derived from their knowledge of typical attacker behavior. For example, a rule could be: "IF a session contains the command \verb|wget| or \verb|curl|, THEN classify the intent as \verb|Malware Deployment|." Our system employs this approach to create a fast and transparent analysis engine.

\section{RELATED WORKS}

Song and firends introduce the Kyoto 2006+ dataset, built from real honeypot traffic over three years, to address the limitations of outdated datasets like KDD'99 in evaluating modern network intrusion detection systems \cite{c1}. Another work \cite{c2} presents the HoneyAnalyzer, a tool that analyzes honeyd log data using an RDBMS and visual interface, allowing precise attack signature extraction through algorithms like "Longest Common Substring". Another study \cite{c3} evaluates honeypots as reliable tools for collecting real attack data, highlighting their advantage over IDSs in avoiding false positives and enabling deeper analysis through automated malware submission. This study focuses on using AI agents efficiently and securely to implement honeypot data analysis.

\section{MODEL}

The model in this study is based on the Cowrie SSH Honeypot. Main functions of Cowrie honeypot are shown in Fig. 1. Cowrie Honeypot is formed of the data collection, emulation and analysis functions. Data Collection: All session logs (commands typed, IP, timestamp) are collected. SSH/Telnet/HTTP are emulated. Extraction of attack patterns is done in the analysis phase. Alert system gives the appropriate warnings.

\begin{figure}[!ht]
\centering
\resizebox{0.48\textwidth}{!}{
\begin{tikzpicture}[
node distance=1cm,
component/.style={rectangle, draw=blue!50, fill=blue!10, thick, minimum width=3cm, minimum height=1cm, rounded corners=0.3cm},
arrow/.style={->, >=stealth, thick}
]

\node[component] (data) {Data Collection (Logs)};
\node[component, below=of data] (analysis) {Analysis Module};
\node[component, left=of analysis] (emulation) {Emulation Layer};
\node[component, right=of analysis] (alert) {Alert System};

\draw[arrow] (emulation) -- (data);
\draw[arrow] (data) -- (analysis);
\draw[arrow] (analysis) -- (alert);

\node[below=0.5cm of analysis, font=\small] {Example: Cowrie SSH honeypot};
\end{tikzpicture}
}
\caption{Cowrie Honeypot Elements}
\label{fig:CowrieElement} 
\end{figure}


\begin{figure}[!ht]
\centering

\begin{tikzpicture}[
    scale=0.70, transform shape,
    node distance=1.5cm,
    component/.style={
        rectangle, draw=blue!50, fill=blue!10, thick,
        minimum width=3cm, minimum height=1.2cm,
        rounded corners=0.3cm, align=center
    },
    db/.style={
        draw=green!50, fill=green!10, thick,
        path picture={
            \draw[green!50] (path picture bounding box.center) ellipse (0.6cm and 0.2cm);
            \draw[green!50] (path picture bounding box.center) ellipse (0.5cm and 0.15cm);
        },
        minimum width=1.5cm, minimum height=1.2cm,
        align=center
    },
    cloud/.style={
        ellipse, draw=orange!50, fill=orange!10,
        minimum width=2.5cm, minimum height=1.5cm, align=center
    },
    arrow/.style={->, >=stealth, thick},
    attack/.style={decorate, decoration={snake, amplitude=1pt, segment length=10pt}, red!80, thick},
    every node/.style={font=\scriptsize}
]

\node[cloud] (attacker) {Attackers};

\node[component, below=of attacker] (frontend) {Frontend \\ (Emulation Layer) \\ {\footnotesize SSH/HTTP/Telnet}};
\node[component, below=of frontend] (logging) {Log Collection \\ {\footnotesize All interaction records}};
\node[db, right=of logging] (database) {Database \\ {\footnotesize MySQL/Elasticsearch}};
\node[component, below=of logging] (analysis) {Analysis Module \\ {\footnotesize AI/Signature Detection}};
\node[component, left=of analysis] (alert) {Alert System \\ {\footnotesize Email/SMS}};
\node[cloud, below=of analysis] (admin) {Admin Interface};

\draw[attack] (attacker) -- node[midway, right] {Attack Attempts} (frontend);
\draw[arrow] (frontend) -- node[midway, right] {Session Logs} (logging);
\draw[arrow] (logging) -- node[midway, right] {Raw Data} (analysis);
\draw[arrow] (logging) -- node[midway, above] {Storage} (database);
\draw[arrow] (analysis) -- node[midway, above] {Anomaly Detection} (alert);
\draw[arrow] (analysis) -- node[midway, right] {Reports} (admin);
\draw[arrow, dashed] (database) |- node[near start, above] {Query} (analysis);

\node[
    draw=gray!50, fill=gray!10, rounded corners,
    text width=8cm, inner sep=6pt, align=left,
    above right=0.5cm and -3cm of frontend
] (desc) {%
\textbf{Typical Honeypot Components:}\\[2pt]
\textbf{1. Emulation Layer}: Fake services that the attacker interacts with.\\[2pt]
\textbf{2. Log Collection}: Recording of all commands, IPs, and timestamps.\\[2pt]
\textbf{3. Analysis Module}: Anomaly detection and extraction of attack patterns.\\[2pt]
\textbf{4. Alert System}: Real-time notifications.%
};

\end{tikzpicture}

\caption{Honeypot Architecture}
\label{fig:HoneypotArchitecture}
\end{figure}
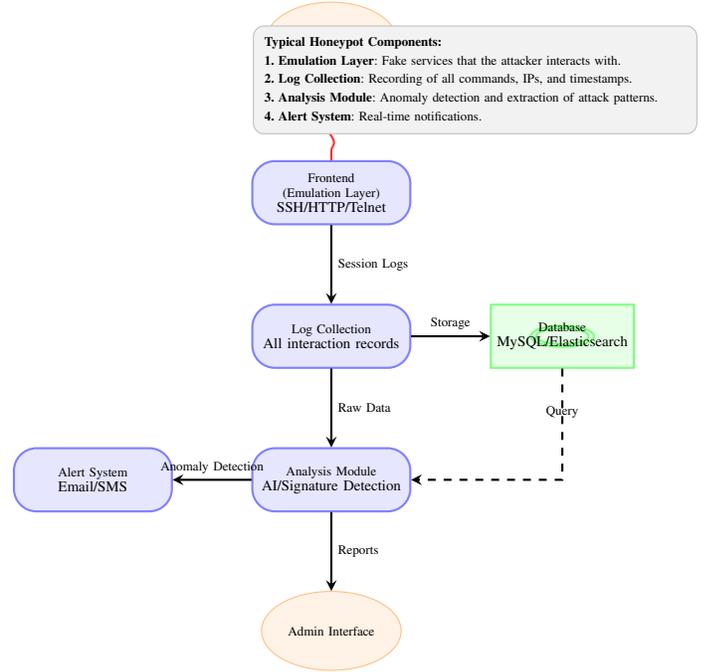


In the context of our project, agentic architecture is deployed with two primary objectives: autonomous system hardening and automated attack pattern (TTP) extraction. The agent utilizes advanced NLP capabilities to not only comprehend the attacker's intent in real-time but to engage them in a dynamic, multi-step dialogue. The agent autonomously and strategically deploys decoy services and credentials to probe the attacker's tools and techniques. Crucially, every command executed, API call made, and lateral movement attempted by the adversary is perceived by the agent as a state within its environment. The agent's planning module then reasons about these states, determining which subsequent deceptive action will best expose the next step in the attack chain. Its integrated memory module continuously logs all interaction sequences in a structured format, creating a rich, contextual dataset of the attack. Furthermore, the agent autonomously executes its core function: pattern extraction. With its tool-calling capabilities, it invokes external analytics modules to process the captured data, clustering similar behaviors, and determining the underlying tactics. This transforms raw log data to a structured and labeled dataset, achieving our ultimate goal. This proactive, agent-driven approach allows us to move beyond mere intrusion detection to active threat modeling and system fortification based on the learned attack patterns.

Our honeypot architecture based on Cowrie is shown in Fig 2. Steps of the model is as follows:
\begin{itemize}
    \item Ingestion: The script begins by discovering and reading all \verb|cowrie*.json| log files from a specified \verb|logs| directory. It handles potential file errors and consolidates all events into a single data structure.
    \item Processing: Using the \verb|pandas| library, the raw list of events is converted into a structured DataFrame. The data is then filtered to isolate command-input events and grouped by the unique \verb|session| ID provided by Cowrie. This step transforms thousands of individual log lines into coherent attacker sessions, each containing the source IP and a chronological list of executed commands.
    \item Analysis: This is the core of the system. A function named \verb|analyze_session_with_rules| iterates through the commands of each session. It uses a predefined dictionary of keywords (e.g., \verb|wget|, \verb|ls|, \verb|rm|) to score each session across several categories (Reconnaissance, Malware Deployment, etc.). Based on these scores, it applies a set of heuristics to classify the session's primary intent and the attacker's estimated skill level. It also extracts suspicious artifacts like URLs for the final report.
    \item Reporting \& Visualization: The results of the analysis are first compiled into a final DataFrame. This DataFrame is then used to generate the final outputs: a \verb|.csv| file for data portability, a \verb|.html| file for easy viewing, and two \verb|.png| image files containing summary visualizations (a bar chart for intents and a pie chart for skill levels).
\end{itemize}

\section{IMPLEMENTATION}

The proposed system was implemented entirely in Python 3, relying on a few key open-source libraries. Pandas is used for all data manipulation tasks, including reading, filtering, grouping, and structuring the data. Matplotlib \& seaborn is used for generating the final data visualizations.

The prototype system is performed on a standard machine (4 CPU cores, 8 GB RAM) and successfully processed over 300,000 log events in a few minutes. 

This project is published on its project GitHub page (\url {https://github.com/EfeEmirYuce/Cowrie-Honeypot-Log-Analysis-Engine}).

The system was evaluated using a dataset of 313,412 log events collected from 10 log files. The pipeline successfully identified and grouped 26,368 unique attacker sessions that contained command inputs. The rule-based engine was able to assign a classification to the vast majority of these sessions, demonstrating the system's ability to handle a significant volume of data and produce a high-level summary.

\section{RESULTS and DISCUSSION}

The application of the analysis process to the dataset yielded insightful results on the nature of attacks targeting the honeypot. The primary finding, as visualized in the generated bar chart, is that the overwhelming majority of attacker sessions were classified with the intent of "Shallow Probe" representing bots or attackers who perform minimal reconnaissance (1-2 commands) before disconnecting. The second most common intent was "Malware Deployment". This indicates that a large amount of attacks are automated and aim to quickly compromise a system to deploy a payload, which is often a cryptocurrency miner or a DDoS bot (see Figure 3). This is further supported by the skill level analysis, which shows that almost all activity is classified as "Low (Script Kiddie)" or "Medium (Automated Script)". This suggests that the honeypot is primarily attracting non-targeted, opportunistic attacks rather than sophisticated, manual operations by skilled adversaries.

The system also generates reports that provide a detailed view of each session, including the source IP and any suspicious URLs extracted, allowing for a deeper dive into specific attack instances.

\begin{figure}
    \centering
    \includegraphics[width=1\linewidth]{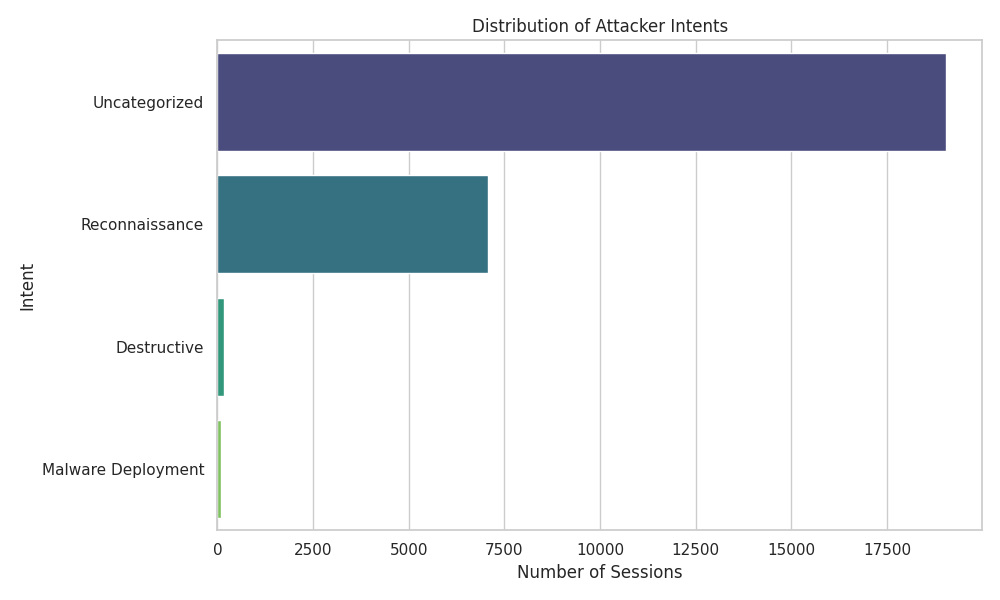}
    \caption{Attack Intent Distribution}
\end{figure}

\section{CONCLUSION }
The analysis of our dataset confirmed that the threat landscape for exposed SSH servers is dominated by automated, low-skill attempts to deploy malware.

This study successfully demonstrated the design and implementation of an automated process for analyzing Cowrie honeypot data. The system effectively transforms high-volume, low-level log data into structured reports and easy-to-understand visualizations. The key contribution is a lightweight, transparent, and extensible process that can be used to derive meaningful threat intelligence. This can be easily setup at honeypot deployments without the need for complex infrastructure. 

For a future work, several enhancements are planned. The rule-based engine could be made more sophisticated by incorporating more complex TTP signatures. Most significantly, the system can be integrated with a Large Language Model (e.g., Google Gemini) via a framework like LangChain. The deployed agents could be equipped with tools to perform real-time threat intelligence enrichment by querying APIs like AbuseIPDB for IP reputation or VirusTotal for URL analysis, providing a much deeper level of contextual understanding for each attack.

\vspace{12pt}

\end{document}